\documentclass[12pt]{article}
\usepackage{latexsym}
\begin{document}

\begin{center}
{\Large \bf Fast Transition between High-soft and Low-soft States in
GRS~1915$+$105: Evidence for a Critically Viscous Accretion Flow\\}
\vspace*{1.2cm}
{\large S. Naik$^1$, A. R. Rao$^1$, and Sandip K. Chakrabarti$^2$\\
}
\end{center}
$^1$ Tata Institute of Fundamental Research, Homi Bhabha Road, Mumbai, India 400~005\\
$^2$ S.N. Bose National Center for Basic Sciences, Salt Lake, Calcutta, 700 091, India\\
\begin{abstract}
We present the results of a detailed analysis of RXTE observations of class
$\omega$ (Klein-Wolt et al. 2002) which show an unusual state transition 
between high-soft and low-soft states in the Galactic microquasar 
GRS~1915$+$105. Out of about 600 pointed RXTE observations, the source was 
found to exhibit such state transition only on 16 occasions. 
An examination  of the RXTE/ASM data in conjunction with the pointed 
observations reveals that these events appeared as a series of quasi-regular 
dips in two stretches of long duration (about 20 days during each occasions) 
when hard X-ray and radio flux were very low. The X-ray light curve and 
colour-colour diagram of the source during these observations are found 
to be different from any reported so far. The duration of these dips is 
found to be of the order of a few tens of seconds with a repetition time 
of a few hundred seconds. The transition between these dips and non-dips 
which differ in intensity by a factor of $\sim$ 3.5, is observed to be 
very fast ($\sim$ a few seconds). It is observed that the low-frequency 
narrow QPOs are absent in the power density spectrum (PDS) of the dip and 
non-dip regions of class $\omega$ and the PDS is a power law in 0.1 $-$ 10 
Hz frequency range. There is a remarkable similarity in the spectral and 
timing properties of the source during the dip and non-dip regions in these 
set of observations. These properties of the source are distinctly different 
from those seen in the observations of other classes. This indicates that 
the basic accretion disk structure during both dip and non-dip regions of 
class $\omega$ is similar, but differ only in intensity. To explain these 
observations, we invoke a model in which the viscosity is very close to 
critical viscosity and the shock wave is weak or absent.
\end{abstract}

keywords: {accretion, accretion discs --- binaries: close ---
black hole physics --- stars: individual: GRS~1915$+$105  ---
X-rays:  stars}

\section{Introduction}
The Galactic microquasar GRS~1915$+$105 was discovered in 1992 with the
WATCH instrument on-board the GRANAT satellite (Castro-Tirado, Brandt, 
\& Lund 1992). Subsequent radio observations led to the identification 
of a superluminal radio source at a distance of 12.5 $\pm$ 1.5 kpc ejecting 
plasma clouds at $v$ $\sim$ 0.92 c (Mirabel \& Rodriguez 1994). Since the 
discovery, the source has been very bright in X-rays, emitting at a 
luminosity of more than 10$^{39}$ erg s$^{-1}$ for extended periods. 
It exhibits peculiar types of X-ray variability characteristics
(Greiner et al. 1996) which have
been interpreted as the instabilities in the inner accretion disk leading 
to the infall of matter into the compact object (Belloni et al. 1997). 
Strong variability is observed in X-ray, radio and infrared over a wide 
range of time scales. Observations with the Burst and Transient 
Source Experiment (BATSE) on the {\it Compton Gamma Ray Observatory} have 
revealed the highly variable nature of the source in the hard X-rays. The 
intensity variations of as much as 3 Crab have been observed on time scales 
from seconds to days (Muno et al. 1999). The X-ray emission is characterized 
by quasi-periodic oscillations (QPOs) at centroid frequencies  in the range 
of 0.001 $-$ 64 Hz (Morgan, Remillard, \& Greiner 1997). It is found that the 
intensity dependent narrow QPOs are a characteristic property of the low-hard
state (Chen, Swank \& Taam 1997). Based on extensive X-ray studies, Muno, 
Morgan, \& Remillard (1999) classified the behavior of the source into two 
distinct states, spectrally hard-state with the presence of narrow QPOs, 
dominated by a power-law component and the soft-state with the absence of 
QPOs, dominated by thermal emission. Attempts have been made to connect 
the observed radio characteristics in GRS~1915$+$105 like jets and 
superluminal motion with the X-ray emission from the accretion disk 
(Naik \& Rao 2000; Naik et al. 2001 and references therein). They have
interpreted the observed soft dips in the X-ray light curves of the source 
as the evacuation of matter from the accretion disk and superposition of a 
series of such dip events produce huge radio flares in the source.

An extensive study of all the publicly available RXTE pointed
observations from 1996 January to 1997 December lead to a classification
into 12 different classes on the basis of structure of the X-ray light 
curve and the nature of the colour-colour diagram (Belloni et al. 2000). 
According to this classification, the source variability is restricted 
into three basic states, a low-hard state with invisible inner accretion 
disk (C), a high-soft state with visible inner accretion disk (B) and a 
low-soft state with spectrum similar to the high-soft state and with much 
less intensity (A). Belloni et al. (2000) suggested that the state 
transition between two canonical states (B and C) takes place through 
a hint of state A. During the transition between low-hard (C) and high-soft 
(B) states, the source shows the properties of state A for a duration of 
about a few seconds. Longer duration ($\geq$ 20 s) of state A (soft-dip) 
has been observed during the transition from low-hard state to high-soft 
state in class $\beta$ (after the spike in the X-ray light curve) and in 
the observations of class $\theta$ when the source exhibits the properties 
of low-hard state (C) during the non-dip regions. Recently, Klein-Wolt al. 
(2002) have discovered an unusual state transition between high-soft and 
low-soft states in the microquasar GRS~1915$+$105. This type of transition 
between two different intensity states was not observed in any other black 
hole binaries.  The low-soft state (state A) is a rare occurrence in Galactic 
Black hole sources. In GRS~1915$+$105 it appears briefly (for a few seconds) 
during the rapid state transition between states C and B as well as during 
the soft dips seen during the variability classes $\beta$ and $\theta$ (in the
nomenclature of Belloni et al. 2000), which are associated with high radio
emission (Mirabel et al. 1998; Naik \& Rao 2000; Naik et al. 2001). On rare
occasions, long stretches of state A are also seen in this source (the
variability class $\phi$). Recently, Smith et al. (2001) reported a sudden
transition from a spectrally hard state to soft state with much lower
intensity in GRS~1758$-$258. Though transition from low-hard to high-soft
states are seen in many Galactic black hole candidate sources, a transition
between two different intensity states (high and low) with similar physical
parameters of the accretion disk was not observed in GRS~1915$+$105 or in
any other black hole binaries.

The extremely variable nature of the microquasar GRS~1915$+$105 is 
restricted within three canonical spectral states A, B, and C. (as 
described above). The source is observed to be in spectral state C (low-hard
state) for wide time ranges starting from hundreds of seconds to tens of days 
to a few months whereas the source remains in the  spectral state B (high-soft 
state) only for a few occasions (Rao et al 2000). These properties are also
seen in other Galactic black hole binaries. The broad-band spectra of the
source obtained from the observations with the Oriented Scintillation 
Spectroscopy Experiment (OSSE) aboard the {\it Compton Gamma Ray Observatory} 
along with the simultaneous observations with RXTE/PCA during lowest X-ray
fluxes (low-hard state C; 1997 May 14$-$20) and highest X-ray fluxes (high-soft
state B; 1999 April 21$-$27) are well fitted by Comptonization of disk 
blackbody photons in a plasma with both electron heating and acceleration 
(Zdziarski et al. 2001). Although the RXTE/PCA observation during 1999 April 
21$-$27 is of class $\gamma$ which is not a pure high-soft state B rather
a combination of states A and B, the overall spectrum is dominated by the
high-soft state B. During the above period, the hard X-ray photon flux 
in 20~$-$~60 keV energy range with BATSE is also found to be very low 
($\sim$ 0.03 photons cm$^{-2}$ s$^{-1}$) which indicates the source 
spectrum to be soft. On a careful analysis of the RXTE/PCA observations
which show high frequency QPO with constant centroid frequency of 67 Hz
,arising in the inner accretion disk of the black hole binary (Morgan et al.
1997), it is found that these observations are of classes $\lambda$, $\mu$, 
$\gamma$. $\delta$ with spectral state of high-soft state B. As these properties
are associated with the inner accretion disk of the black hole, it is interesting
to study the RXTE/PCA observations during the high-soft states with low value of
hard X-ray photon flux with BATSE in detail.

In this paper, we present the evidence of fast transitions 
between two different X-ray intensity states with similar spectral and timing 
properties in GRS~1915$+$105, during the high-soft state of the source
(class $\omega$). A detailed spectral and timing analysis is presented
which show that the low-soft state is very different from the spectral state
A seen during other variability classes.
We have tried to explain the observed peculiar 
state transition in GRS~1915$+$105 on the 
basis of the presence of an accretion disk with critical viscosity which causes 
the appearance and disappearance of sub-Keplerian flows out of Keplerian matter.

\section {Analysis and Results}

\setcounter{figure}{0}
\begin{figure*}
\vskip 6.0cm
\includegraphics{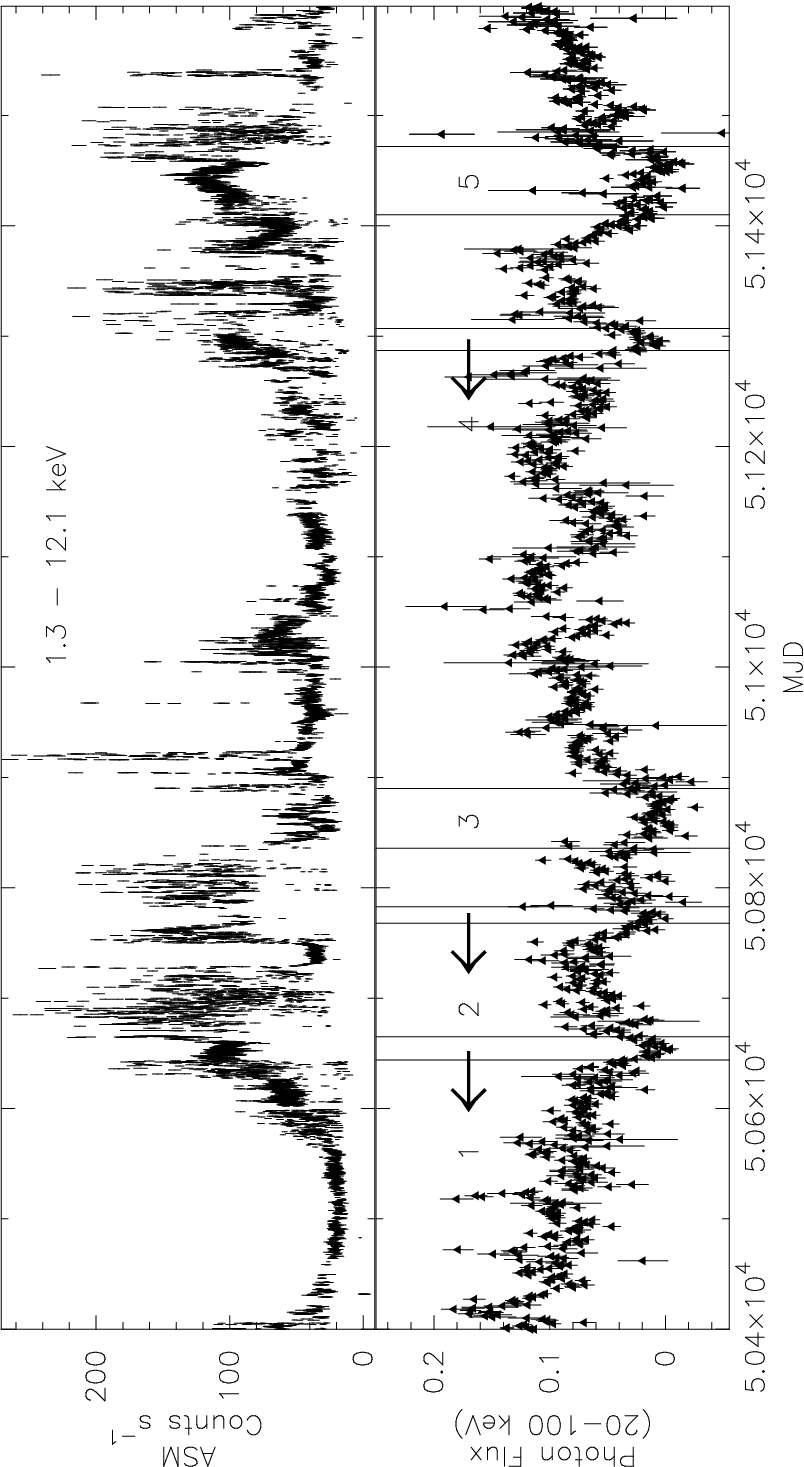}
\caption{The X-ray light curve for GRS 1915$+$105 with RXTE/ASM in the energy
range 1.3$-$12.2 keV is shown in the upper panel with 
the hard X-ray photon flux in the energy 
range 20~$-$~60 keV with BATSE in the bottom panel. The regions (1), (2), (3), 
(4), and (5) in the bottom panel indicate the presence of the long duration 
($\geq$ 10 days) soft-spectral states of the source when the hard X-ray 
photon flux of the source is $\sim$ 0.03 photons cm$^{-2}$ s$^{-1}$.}
\end{figure*}

We have made a detailed examination of all the publicly available RXTE 
pointed observations on GRS~1915$+$105 in conjunction with the continuous 
monitoring of the source using RXTE/ASM. Based on the X-ray light curve
and the hardness ratio, we could identify most of the pointed observations
into the 12 variability classes suggested by Belloni et al. (2000), and
associate the global properties of the source with other characteristics
like radio emission (see Naik \& Rao 2000). During these investigations
a new variability class was found to be occurring during two time intervals, 
1999 April 23 $-$ May 08 (MJD 51291$-$50306) and 1999 August 23 $-$ September 
11 (MJD 51410$-$51432), respectively. This new class which is called as 
class $\omega$ (Klein-Wolt et al. 2002), was observed in a total of 16 
pointed RXTE observations and is characterized by a series of dips of 
duration of 20$-$95 s and repetition rate of 200$-$600 s.

\setcounter{figure}{1}
\begin{figure*}
\vskip 6.0cm
\includegraphics{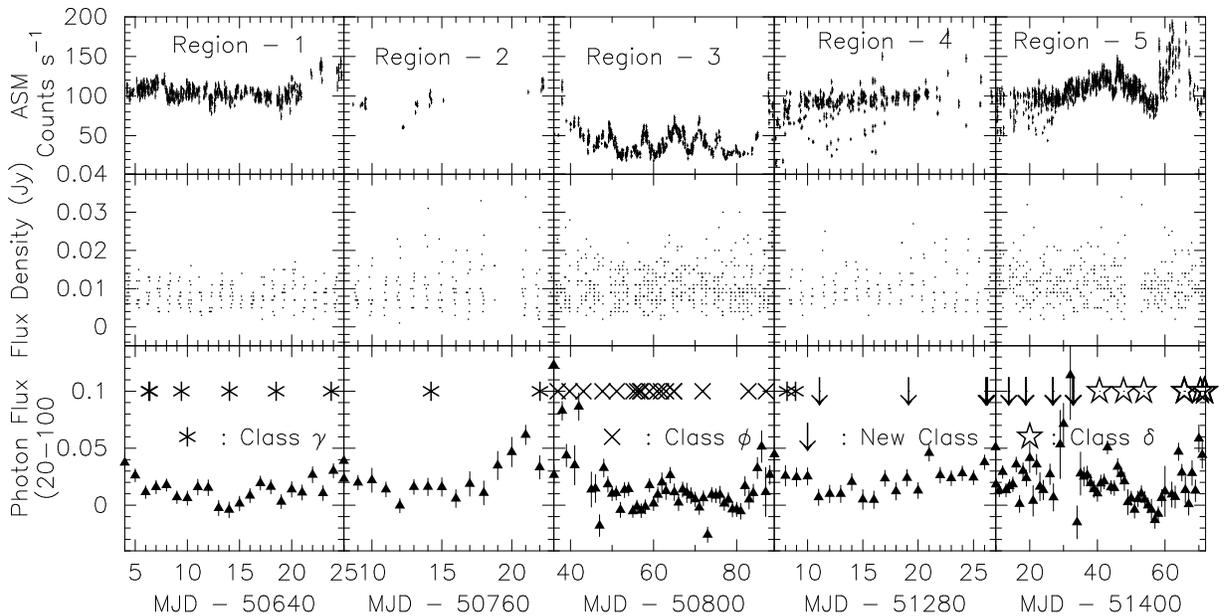}
\caption{The X-ray light curve for GRS 1915$+$105 with RXTE/ASM in the energy
range 1.3$-$12.2 keV during the 5 different regions marked in Figure 1 
(bottom panel) is shown in the upper panel with the 
radio flux at 2.25 GHz (second panel) and hard X-ray photon 
flux in the energy range 20 $-$ 60 keV with BATSE (bottom panel). The 
start time of RXTE pointed observations during these five intervals are
indicated by different markers in the bottom panel of the figure.}
\end{figure*}

To compare and contrast the X-ray properties of the  source during this new
class with other reported classes, we show, in Figure 1, the X-ray light 
curve of the source obtained with RXTE/ASM in 1.3~$-$~12.1 keV energy range 
(top panel) with the hard 
X-rays photon flux in 20~$-$~60 keV energy range (bottom panel). The data for 
hard X-rays were taken from the Burst and Transient Source Experiment (BATSE) 
on the {\it Compton Gamma Ray Observatory}. We have selected the time range 
(MJD~50400~$-$~51600) during which the X-ray (1.3~$-$~12.1 keV), radio and 
hard X-ray (20~$-$~60 keV) data are available. We have selected five regions 
of durations of more than about 10 days when the hard X-ray photon flux is 
$\sim$ 0.03 photons cm$^{-2}$ s$^{-1}$. These regions are marked by vertical 
lines in the bottom panel of Figure 1.

To examine the variation in the source flux at different energy bands
(X-ray, radio, and hard X-ray photon flux), we have plotted in Figure 2,
the ASM light curve in 1.3$-$12.2 keV energy range (top panel),  
radio flux density at 2.25 GHz (middle panel), and 
the hard X-ray photon flux (bottom panel) during above five different regions.
The start time of RXTE pointed observations during these five intervals and 
the class (Belloni et al. 2000) to which these observations belong to, are 
marked in the bottom panel of the figure. The average ASM count rate, average flux density at 2.25 GHz, spectral 
index, the hard energy photon flux and the rms variation in above 
parameters during all these five regions are given Table 1. From the table,
it is observed that average ASM count rate during first, second and fifth
regions are $\sim$ 100 counts s$^{-1}$ and $\sim$ 80 counts s$^{-1}$ 
during fourth region whereas the count rate is too low ($\sim$ 40 counts 
s$^{-1}$) during the third region. It is observed that the rms variation
in the source count rate is maximum during the third region. 
However, the radio 
flux density at 2.25 GHz and the hard X-ray photon flux are found to be 
indifferent during all these five intervals. From a careful analysis of the 
RXTE pointed observations during these five intervals, it is found that the 
RXTE observations are of class $\gamma$ (regions 1, 2, and two observations in 
region 3), class $\phi$ (region 3), class $\delta$ (half of the region 5) and 
class $\omega$ (new class; regions 4 and 5). These observations and the classes
are indicated in the bottom panel of Figure 2. The X-ray light curves of the 
RXTE pointed observations in 1999 April~$-$~May and 1999 August~$-$~September 
(class $\omega$) which show a quasi-regular and distinct transition between two 
different X-ray intensity states (dip and non-dip) are different from the 
reported 12 different classes. 

\setcounter{table}{0}
\begin{table*}
{\small
\caption{Statistics of five different regions shown in Figure-2}
\begin{center}
\begin{tabular}{llllll}
\hline
\hline
   &Region -- 1  &Region -- 2 &Region -- 3 &Region -- 4 &Region -- 5\\
\hline
\hline
ASM Count rate$^1$  &103.4$\pm$2.8  &91.3$\pm$6.7   &37.6$\pm$7.4  &78.9$\pm$11.7   &102.8$\pm$6.5\\
Flux density$^2$    &0.009$\pm$0.0002 &0.011$\pm$0.0006 &0.011$\pm$0.0002 &0.011$\pm$0.0005 &0.011$\pm$0.0003\\
Spectral Index  &--0.11$\pm$0.05  &--0.15$\pm$0.06 &--0.25$\pm$0.03 &--0.06$\pm$0.06 &+0.07$\pm$0.04\\
Photon flux$^3$     &0.014$\pm$0.002 &0.022$\pm$0.006  &0.009$\pm$0.003 &0.023$\pm$ 0.003 &0.018$\pm$0.002\\
 \hline
 \hline
\multicolumn{6}{l}{$^1$ The average ASM countrate obatained from the Dwell data and the quoted errors are the}\\
\multicolumn{6}{l}{~~~~~~~~~~~~~~~~ rms deviations from the average countrate.}\\
\multicolumn{6}{l}{$^2$ Flux density in mJy at 2.25 GHz}\\
\multicolumn{6}{l}{$^3$ Hard X-ray flux from BATSE in photons cm$^{-2}$ s$^{-1}$}\\
\end{tabular}
\end{center}
}
\end{table*}
Out of about 600 RXTE pointed observations, there are only 16 occasions 
when the source shows the unusual transition between two different intensity 
states (class $\omega$). We emphasize here that these 16 observations
occur only during two occasions (51290$-$50306 and 51410$-$51433 MJD ranges) 
when the hard X-ray flux was low (as described above). We have shown, in 
Figure 3, the X-ray light curve for one such observation carried out on 1999 
April 23 (Obs. ID: 40403-01-07-00). The panel (a) in the Figure shows the 
RXTE/PCA light curve of the source in 2~$-$~60 keV energy band (normalized to 5 
Proportional Counter Units (PCUs)) whereas the panel (b) and (c) show the 
hardness ratios HR1 (the ratio between the count rate in the energy range 
5$-$13 keV to that in 2$-$5 keV) and HR2 (the ratio between the count rates 
in the energy range 13$-$60 keV and 2$-$13 keV), respectively. In panels (d), 
(e), and (f), we have shown the source light curves (normalized to 5 PCUs) 
in 2$-$5 keV, 5$-$13 keV, and 13$-$60 keV energy bands respectively. The 
average value of source count rate and rms variations in the X-ray flux in 
different energy bands during the dip, non-dip, and the total light curves 
are given in Table 2 along with the hardness ratios HR1 and HR2. From Figure 3
and Table 2, it can be seen that the variability in the source flux is low 
during the dips in all the energy bands. It is, however, observed that the
source variability decreases at high energy bands. An intensity difference 
by a factor of $\geq$ 3 is observed between the non-dip and dip regions at
low energy bands which $\sim$ 2 at hard X-ray bands. This indicates that the
change in the source flux during dip and non-dip regions are significant in soft
X-ray bands which decreases significantly in hard X-ray bands. The low 
intensity dips are characterized by HR1 and HR2 in 0.35~$-$~0.55 and 
0.03~$-$~0.06 ranges respectively whereas the non-dip regions are 
characterized by HR1 and HR2 in 0.6~$-$~0.8 and 0.15~$-$~0.25 ranges 
respectively. The observed differences between the hardness ratios HR1 and 
HR2 during the dip and non-dip regions are not significant enough to highlight 
the difference in the spectral properties of the source.

The duration of the dips in the light curves of all the 16 RXTE pointed 
observations of class $\omega$ which show the 
unusual transition between two different intensity states, lies in the 
time range of 20~$-$~95 s whereas the non-dip regions last for 200~$-$~525 
s. Although the duration of the dips was high in the beginning of the 
observations during both the occasions, sparse RXTE pointed observations 
restrict us  to 
present any statistical picture of the duration and the repetition period 
of these dips.  Figure 3 shows that the HR1 is low during the dips. 

\setcounter{figure}{2}
\begin{figure*}
\vskip 8.6cm
\includegraphics{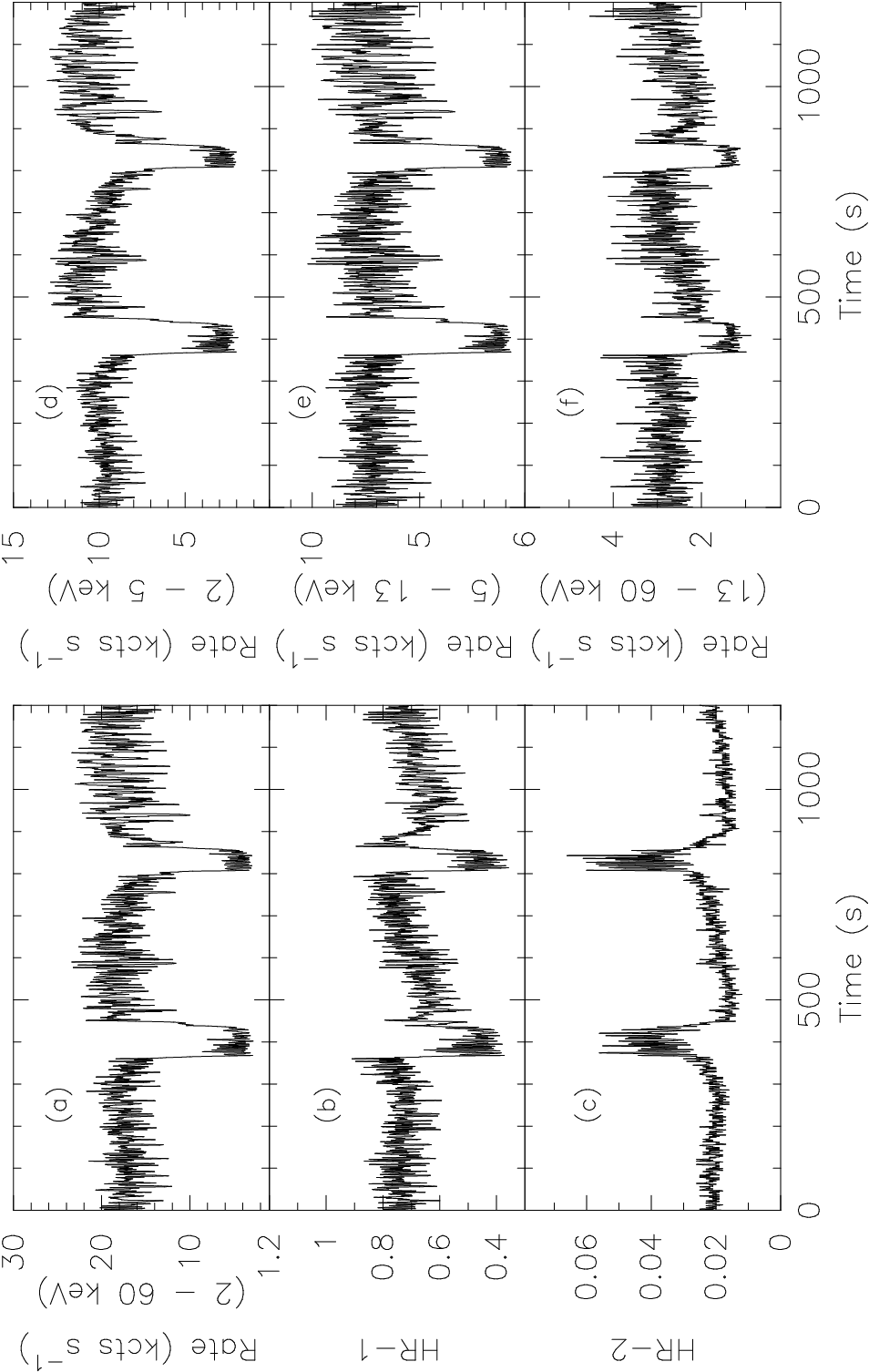}
\caption{Light curve (2~$-$~60 keV energy range for 5 PCUs) of GRS~1915$+$105
for the RXTE/PCA pointed observation on 1999 April 23 (new class $\omega$) is 
shown along with the hardness ratios HR1 (count rate in 5~$-$~13 keV / count 
rate in 2~$-$~5 keV energy range) and HR2 (count rate in 13~$-$~60 keV / 
count rate in 2~$-$~13 keV energy range). The light curves at 2$-$5 keV,
5$-$13 keV, and 13$-$60 keV energy bands are shown in the right panels. 
Low values of HR1 ($\leq$ 1.0) and HR2 ($\leq$ 0.06) indicate the softness 
of the spectrum during both the dip and non-dip regions.}
\end{figure*}

\setcounter{table}{1}
\begin{table}[h]
\caption{Statistics of the non-dip and dip regions shown in Figure-3}
\begin{center}
\begin{tabular}{lllllllllllll}
\hline
\hline
Energy Range   &\multicolumn{3}{|c|}{Average Countrate$^1$}  &Flux\\
(in keV)       &Non-dip Region  &Dip Region  &Total (dip and non-dip) &Ratio$^2$\\
\hline
\hline
~~2 $-$ 5   &9744$\pm$46   &3171$\pm$95  &8991$\pm$68   &3.1$\pm$0.1 \\
~~5 $-$ 13  &7147$\pm$55   &1593$\pm$78  &6298$\pm$57   &4.5$\pm$0.22\\
13 $-$ 60   &2787$\pm$23   &1393$\pm$29  &2511$\pm$17   &2.0$\pm$0.05\\
~~2 $-$ 60  &17239$\pm$101 &4938$\pm$175 &15603$\pm$125 &3.5$\pm$0.13\\
 \hline
 \hline
\multicolumn{8}{l}{$^1$ Average Count rate (in Counts s$^{-1}$)}\\
\multicolumn{8}{l}{$^2$ Ratio between the average count rate during the non-dip region to the dip region} \\
\end{tabular}
\end{center}
\end{table}

To study the timing properties of the source, we have generated the power 
density spectrum (PDS) in 2~$-$~13.1 keV energy band for the dip (low intensity
state) and non-dip (high intensity state) regions of all 16 selected RXTE/PCA 
pointed observations. It is found that the low frequency narrow QPOs are absent
in the PDS of both dip and non-dip regions of all the observations. 
We have fitted the PDS of both the dips and non-dip regions with a power-law 
in frequency ranges 0.1~$-$~1 Hz and 1~$-$~10 Hz. It is found that 
there is no significant difference in the power law index during both the 
intensity states. The only distinguishing feature of the two intensity states 
is the higher rms variability in both the frequency bands during the low 
intensity (dip) state compared to the high intensity (non-dip) state. Figure 4 
shows the PDS of the two different regions of the RXTE/PCA pointed observation 
on 1999 April 23. From the Figure, it is found that the PDS during both the 
regions are a featureless power law in 0.1~$-$~10 Hz range.

\setcounter{figure}{3}
\begin{figure*}
\vskip 8.4cm
\includegraphics{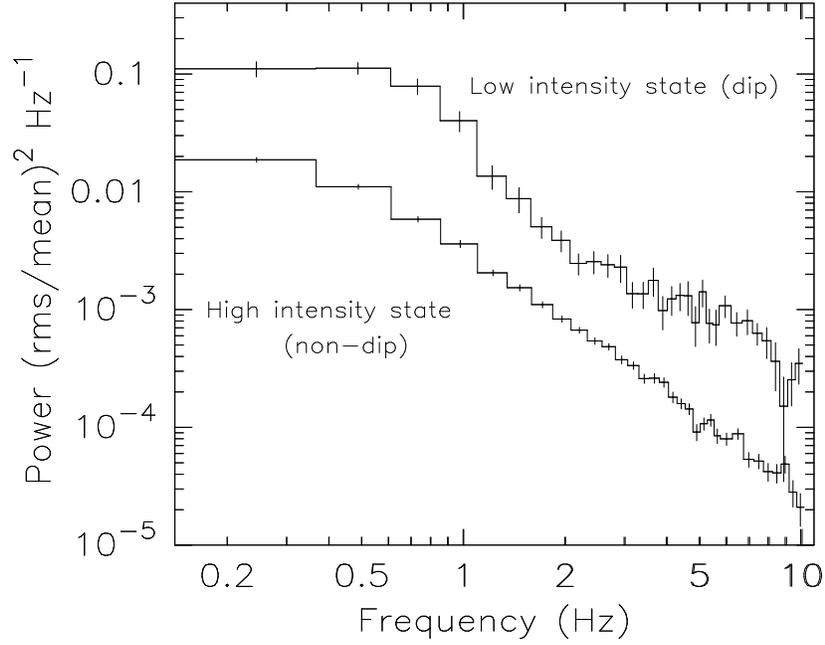}
\caption{PDS of the source GRS 1915$+$105 in the energy range of 2 $-$ 13 keV
for high-soft and low-soft states of the X-ray light curve obtained from
RXTE/PCA on 1999 April 23. The absence of the narrow QPO in the frequency
range of 0.1 $-$ 10 Hz is clear. The normalized power at a given frequency
is high for the low-soft (dip) state and is low for the high-soft state
(non-dip).}
\end{figure*}

\setcounter{figure}{4}
\begin{figure*}
\vskip 8.5cm
\includegraphics{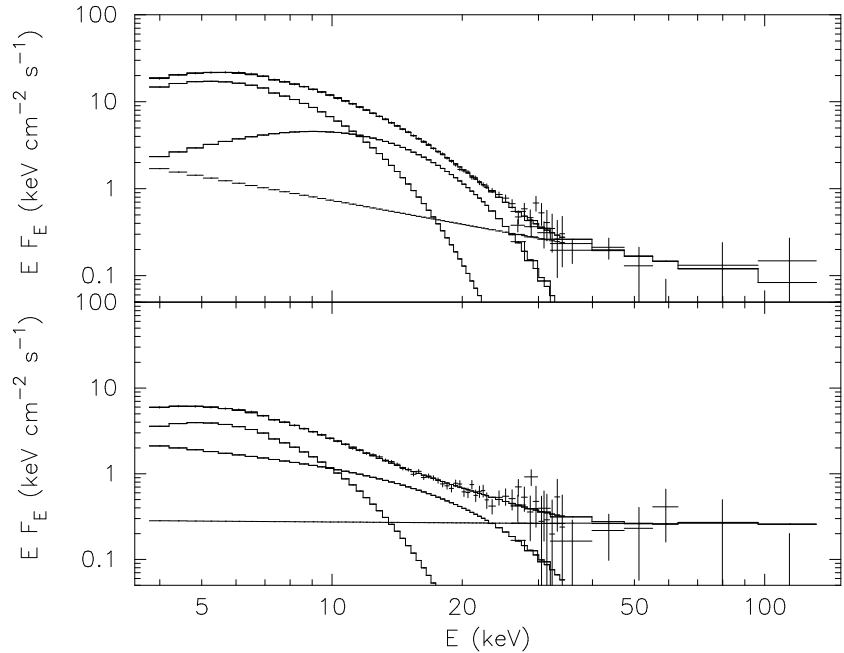}
\caption{The observed count rate spectrum of GRS~1915+105 during high-soft
(non-dip) and low-soft (dip) states of the new class $\omega$ obtained 
from RXTE/PCA and HEXTE data. A best-fit model consisting of a disk 
blackbody, a power-law, and a thermal Compton spectrum is shown as histogram 
with the data. }
\end{figure*}

We have attempted a wide band X-ray spectroscopy of the dip and non-dip
regions of the RXTE observation of the source on 1999 August 23 (Obs. ID :
40703-01-27-00). The data for the dip region were selected when the source 
count rate was less than $\leq$ 3000 counts s$^{-1}$ (for 2 PCUs) in 2$-$60 
keV energy range and non-dip region when the count rate was $\geq$ 5000 counts 
s$^{-1}$. We have generated 128 channel energy spectra from the standard 2 
mode of the PCA and 64 channel spectra from HEXTE for the dip and non-dip 
regions. Standard procedures for data selection, background estimation and 
response matrix generation have been applied. Systematic error of 2\% have 
been added to the PCA spectral data. We have used the archive mode data from 
Cluster 0 of HEXTE for better spectral response. The spectra were re-binned at 
higher energy band to fewer number of channels in order to improve the 
statistics. 3~$-$~50 keV energy range PCA data and 15~$-$~180 keV energy range 
HEXTE data are used for spectral fitting. The dip and non-dip spectra are 
fitted with the standard black hole models (Muno et al. 1999) consisting of 
``disk-blackbody and a thermal-Compton spectrum'', ``disk-blackbody and a 
power-law'', and ``disk-blackbody, a power-law and a thermal-Compton spectrum''
with a fixed value of absorption by intervening cold material parameterized as 
equivalent Hydrogen column density, N$_H$ at 6 $\times$ 10$^{22}$ cm$^{-2}$. 
From the spectral fitting, it is observed that the model with disk-blackbody, 
a power-law, and a thermal-Compton spectrum as model components fits very well 
with the data during both the dip and non-dip regions. It is observed that
the source spectrum is similar in hard X-ray energy bands ($\geq$ 50 keV)
for both the dip and non-dip regions. The fitted parameters 
for the best fit model during two different regions along with the 2$-$50 keV
source flux for each model components are given in Table 3. Assuming the 
distance of the source as 12.5 kpc, we have calculated the luminosity of the 
source in 3~$-$~60 keV energy band to be $\sim$ 7.83 $\times$ 10$^{-38}$ 
ergs s$^{-1}$ and 2.2 $\times$ 10$^{-38}$ ergs s$^{-1}$ for non-dip and dip 
regions respectively. The parameters in the table show the similarities
in the properties of the accretion disk during two different intensity states. 
From the results of the spectral fitting, we found that the spectrum of the 
source is soft with similar parameters of the accretion disk during both the 
different intensity states. We have shown, in Figure 5, the energy spectra 
obtained from the RXTE/PCA and HEXTE observations of the source with 
the fitted model (``disk-blackbody, a power law and a thermal Compton spectrum)
during two different intensity states. Upper panel in Figure 5 shows the 
the energy spectrum and the best fit model for the non-dip region of class 
$\omega$ whereas the bottom panel shows the spectrum and the fitted model
for the dip regions. From the Figure, it is observed that the dip and non-dip 
spectra are dominated by the thermal component. 

In order to investigate the structure of the inner accretion disk during
the observed unusual transition between two different intensity states,
we have calculated the characteristic radius of the inner disk
($R_{col} = D_{10kpc}\sqrt{N_{bb}/cos\theta}$) from the normalization
parameter of the disk blackbody (N$_{bb}$). Assuming the distance of the
source to be 12.5 kpc (D$_{10kpc}$ = 1.25), and an inclination angle
($\theta$) equal to that of the radio jets, 70$^{\circ}$, the radius of
the inner accretion disk is found to be 41$\pm$7 km during the non-dip
(high intensity state) and 25$\pm$3 km during the dip (low intensity state)
regions. The temperature of the inner accretion disk during the dip and
non-dip regions are found to be 1.54 and 1.72 keV respectively. Using these
values, we have estimated the ratio of the total flux from the disk ($F_{bb}
= 1.08 \times 10^{11}N_{bb}\sigma T_{col}^4$ ergs$^{-1}$cm$^{-2}$s, where
$\sigma$ is the Stephan-Boltzmann constant) during non-dip and dip regions
to be $\sim$ 3.9 which is observed from the X-ray light curves.
From this analysis, we conclude that the observed transition between
the different intensity states are associated with the change in temperature
of the inner accretion disk without any change in the radius.

\setcounter{figure}{5}
\begin{figure*}
\vskip 7.2cm
\includegraphics{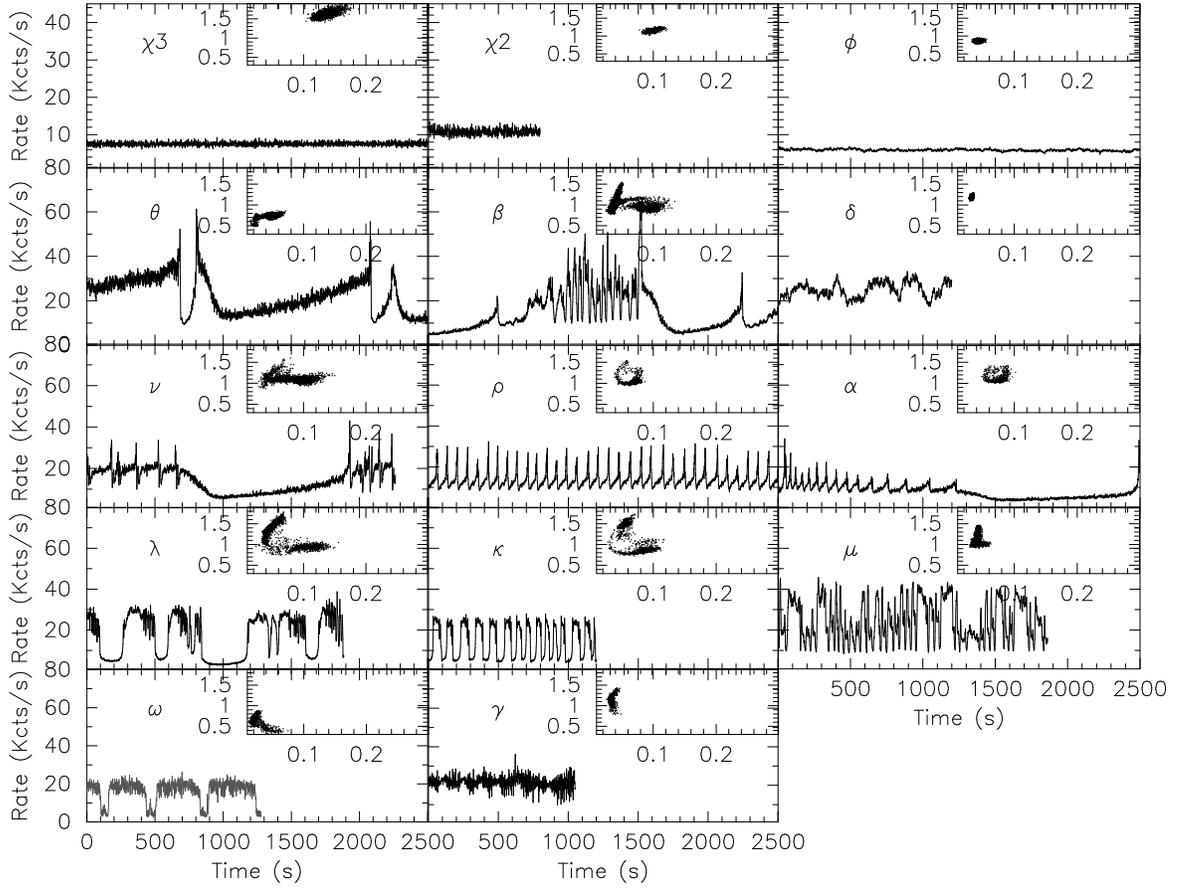}
\caption{X-ray light curves (2$-$60 keV energy range) and colour-colour
diagram (HR1 vs HR2) of GRS~1915$+$105 for observations of all classes 
(Belloni et al. 2000) are shown along with the observation of class $\omega$.
The insets in each figure shows the colour-colour diagram, 
HR1 in the Y-axes and HR2 in the X-axes (see text). The Obs. Ids of the data
used are $\chi3$ : 20402-01-50-01, $\chi2$ : 20402-01-04-00, $\phi$ : 
30703-01-08-00, $\theta$ : 40702-01-03-00, $\beta$ : 20402-01-44-00, $\delta$ :
20402-01-42-00, $\nu$ : 10408-01-41-00, $\rho$ : 20402-01-03-00, $\alpha$ : 
20402-01-28-00, $\lambda$ : 20402-01-37-01, $\kappa$ : 20402-01-33-00, $\mu$ :
20402-01-45-01, $\omega$ : 40403-01-07-00, and $\gamma$ : 20402-01-40-00.}
\end{figure*}

\section{Comparison between various spectral states}

The spectral and temporal properties of the source were studied by Rao et al. 
(2000) when the source was making a slow transition from a low-hard state 
(C) to a high-soft state (B) in about 3 months. Rapid state transitions  
between above two canonical spectral states were also observed when the 
source was exhibiting a series of fast variations, which can be classified 
as bursts (Rao et al. 2000). It was pointed that the spectral and timing
properties of the source during the short duration B and C states are 
identical to those seen during the long duration B and C states.
Fast transition between low-hard (C) and low-soft
states (A) were also observed when the light curve of the source contains
a series of X-ray soft dips (Naik et al. 2001). Although Belloni et al. (2000)
have classified the RXTE/PCA observations of low-hard states into four 
different sub-classes ($\chi1$, $\chi2$, $\chi3$, and $\chi4$), there is 
not much difference in the temporal and spectral properties of the source
during the observations of these four sub-classes. During the observations
of class $\phi$, the source remains in low-soft state (A) whereas during all
other classes, it is observed that the source remains in high-soft spectral 
state (B) or makes transition between the spectral states B and C, and C and 
A. Although the transition between the spectral states B and A is seen during 
the observations of a few classes, the duration of the spectral state A is 
in the order of a few seconds. However, the transition between the states A 
and B as observed during the observations of class $\omega$ where the source 
remains in state A for a few tens of seconds is different. As both the spectral
states A and B are characterized by the soft spectrum with inner accretion disk
extending towards the black hole event horizon, the difference in the observed 
transition between these two states needs a detailed comparison between these 
three states along with the low-soft state observed during the class $\omega$.

According to the Belloni et al. (2000) classification, 
the source variability ranges
from a steady emission for long durations like in classes $\phi$, $\chi$
to large amplitude variations in classes $\lambda$, $\kappa$, $\rho$, $\alpha$.
Short periodic flickering with different amplitudes is seen during the 
observations of classes $\gamma$, $\mu$ and $\delta$. During the observations 
of classes $\theta$, $\beta$, and $\nu$, the amplitude variation is 
accompanied by soft X-ray dips with duration of a few tens of seconds 
to hundreds of seconds. It is observed that the properties of the source
during the observations of class $\alpha$ are similar to those during the 
combined classes $\rho$ and $\chi$ (Naik et al. 2002). However, the RXTE 
pointed observations which show the unusual transition between two different 
intensity states (class $\omega$) are found to be different from the all 
other classes. The absence of strong variability in the X-ray light curve, 
the presence of soft-dips with similar spectral properties as the non-dip 
regions prompted to investigate the similarities/differences in the source 
properties during various low and high intensity states with little variability 
in the X-ray light curve of the reported 12 different classes of RXTE 
observations. 

We have selected one RXTE observation from all the X-ray classes 
and, in Figure 6, we have shown the light curves and colour-colour (HR1 vs HR2)
diagram of the selected observations. From the Figure, it is observed that 
observations of classes $\phi$, $\delta$ and $\omega$ are characterized by the 
absence of strong variability in the light curve with HR1 $\leq$ 1 and HR2 
$\leq$ 0.06 which indicate the softness of the spectrum of the source. Although
the observations of classes $\chi2$ and $\chi3$ also show less variability
in the X-ray light curves, the spectrum of the source is hard with HR1 
$\geq$ 1 and HR2 $\geq$ 0.1. Comparing the structure of the light curves and 
colour-colour diagram of all these observations, it is found that the 
observation which show the unusual intensity transition (class $\omega$) is 
different from the observations of other classes. We have examined the 
uncertainty in the value of HR1 (about 0.3) due to the different gains of 
PCUs at different epochs. But the general trend in the hardness ratios remain 
unaffected by the change in gains.

To study the spectral properties of the source during these different
classes, we have attempted a wide band X-ray spectroscopy of all the RXTE 
observations which show a gradual change from a high-soft to low-hard state 
and characterized by a sharp soft dip with low variability ($\beta$ and 
$\theta$) and observations which show steady behavior during the orbit of 
RXTE ($\chi$2, $\chi$3, $\phi$, and $\delta$) along with one of the 
observation showing the unusual state transition (class $\omega$). 
We have selected the radio-quiet low-hard state of class $\chi2$, radio-loud 
low-hard state of class $\chi3$, and the high-soft (non-dip) state of the 
observations of class $\omega$. Standard procedures for data selection, 
background estimation and response matrix generation have been applied. 
We have used the data in same energy range such as 3~$-$~50 keV energy range 
PCA data and 15~$-$~180 keV energy HEXTE data for spectral fitting. The 
spectra of all three different classes of observations are fitted with all 
the three models which were used for the dip and non-dip regions (described 
in the previous section). The fitted parameters for different models are given 
in Table 4. Examining the parameters in the table, it is found that 
``disk-blackbody and a thermal-Compton spectrum'' model is suitable for the 
radio-quiet low-hard state (class $\chi$2) and ``disk-blackbody, a power-law 
and a thermal-Compton spectrum'' model is suitable for the radio-loud low-hard 
state (class $\chi$3) and the high state (non-dip) of the observations which 
show the unusual transition between two intensity states. 

To compare the deviations of the spectral parameters of the source during 
the various low state observations with the above described three well 
fitted models for the observations of classes $\chi$2, $\chi$3, and the 
non-dip region of the observation showing the peculiar state transition, 
we analyzed the source spectrum during various dips observed in other 
classes of RXTE observations along with a few low variability high-state 
observations of classes $\phi$ and $\delta$. We have selected data for 
the low-hard dip of class $\beta$ (before the spike in the light curve), 
low-soft dip of class $\theta$ which is identical to the dip observed (after 
the spike) in the light curve of class $\beta$, low-intensity observation 
of class $\phi$ and high-soft state of class $\delta$. To investigate the 
similarities in the spectral properties of the source during above X-ray 
observations, we have used ``disk-blackbody and a thermal-Compton spectrum'' 
and ``disk-blackbody, a power-law and a thermal-Compton spectrum'' models to 
fit the spectra. All the spectral parameters other than the normalizations 
are fixed for the models which are well-fitted with the data during the 
radio-quiet low-hard state (class $\chi2$), radio-loud low-hard state 
(class $\chi3$) and the high state of the observations of the class $\omega$. 
The fitted parameters are shown in Table 5. Comparing the fitted parameters
in Table 5, it is observed that the model for the high-soft state of
class $\omega$ fits better with the spectra during the dip of the class $\omega$
(marked as A' in the Table 5), soft-dip of class $\theta$ and the observation 
of class $\delta$ whereas the $\chi$2 model fits better with the spectrum 
of the low-hard dip of class $\beta$. From the table, it is also noticed that
the spectrum of the observation of class $\phi$ does not fit to any of the
models and needs more investigations on the spectral properties of the source. 
We can draw two important conclusions from the spectral
analysis: (1) the low soft state seen during class $\omega$, though
generically can be classified as spectral state A, it is distinctly different
from the long duration spectral state  A (class $\phi$) and short 
duration A states seen during class $\theta$ etc. (2) the spectral shape
during the low soft state is similar to the high soft state (with the
same set of spectral components).
Hence we can conclude that 
the high-soft and
low-soft states of the observations of class $\omega$ do not have any 
significant difference in the physical parameters of the accretion disk of 
the black hole except the normalization factors and hence these two different 
intensity states are identical in the spectral and temporal 
properties of the source and different from the observations of any other 
reported classes.

\setcounter{table}{2}
\begin{table*}
\caption{Spectral parameters during high-soft and low-soft states of class $\omega$ 
of GRS~1915$+$105 for the model ``disk-blackbody, a power law and a thermal Compton 
spectrum''}
\begin{center}
{\small
\begin{tabular}{lllllccc}
\hline
\hline
X-ray intensity  &Reduced   &$kT_{in}^{1}$ &$kT_{e}^{2}$  &$\Gamma_{x}^{3}$ & R$_{in}$  &L$^4$  &\\
state   &$\chi^{2}$  &(keV)  &(keV)   &    &(km)   &ergs s$^{-1}$ & \\
\hline
\hline
High-soft (non-dip) &0.68 (76 dof) &1.7$_{-0.06}^{+0.05}$  &2.5$_{-0.15}^{+0.13}$   &2.9$_{-0.97}^{+0.2}$ &42$_{-3}^{+5}$ &7.83 $\times 10^{-38}$\\
Low-soft (dip)   &0.68 (66 dof) &1.55$_{-0.05}^{+0.03}$   &4.05$_{-0.045}^{+0.042}$  &2.03$_{-0.1}^{+0.1}$ &25$_{-3}^{+6}$ &2.2 $\times 10^{-38}$\\
\hline
\hline
\multicolumn{6}{l}{$^{1}kT_{in}$ : Inner disk temperature,~~$^2kT_{e}$ : Temperature of the Compton cloud,}\\
\multicolumn{6}{l}{$^{3}\Gamma_{x}$ : Power-law photon index,~~$^4$L : Luminosity of the source in 3$-$60 keV}\\
\multicolumn{6}{l}{energy range, assuming the distance of the source to be 12.5 kpc.}\\
\end{tabular}
}
\end{center}
\end{table*}

\setcounter{table}{3}
\begin{table}
\caption{Spectral parameters during classes $\chi2$, $\chi3$, and $\omega$ 
(non-dip) of GRS 1915$+$105}
\begin{center}
\begin{tabular}{llllllll}
\hline
\hline
X-ray            &Reduced     &$kT_{in}^{1}$ &$kT_{e}^{2}$  &$\tau^{3}$ &$\Gamma_{x}^{4}$ &\multicolumn{2}{c}{Count rate} \\
Class            &$\chi^{2}$  &(keV)     &(keV)     &       &   &HEXTE  &PCA\\
\hline
\hline
\multicolumn{8}{c}{Model: Disk blackbody + thermal-Compton spectrum} \\
\hline
\hline
$\chi2$ (RQ$^5$)     &1.353      &1.348     &20.01   &3.047 &$---$  &62.58  &4093\\
$\chi3$ (RL$^6$)	 &8.26       &4.068     &12.54   &4.159 &$---$  &78.22  &4902\\
$\omega$       &2.035      &2.056     &4.415   &6.489 &$---$  &71.08  &8547\\ 
(High state)	 &  &  &  &  &  &  &\\
\hline
\hline
\multicolumn{8}{c}{Model: Disk blackbody + power-law}\\
\hline
\hline
$\chi2$ (RQ)	 &5.313   &0.156    &$---$ &$---$    &2.519  &$---$ &$---$\\
$\chi3$ (RL)	 &33.24   &0.156    &$---$ &$---$    &2.636  &$---$ &$---$\\
$\omega$	 &2.505   &2.205    &$---$ &$---$    &3.457  &$---$ &$---$\\
(High state)  & & & & & & &\\
\hline
\hline
\multicolumn{8}{c}{Model: Disk blackbody + power-law + thermal-Compton spectrum} \\
\hline
\hline
$\chi2$ (RQ)   &1.367   &1.349    &20.01   &3.047  &2.52   &$---$ &$---$\\ 
$\chi3$ (RL)   &1.736   &2.58     &4.464   &46.56  &2.606  &$---$ &$---$\\
$\omega$       &0.68    &1.71     &2.532   &25.42  &2.91   &$---$ &$---$\\
(High state)  & & & & & & &\\
\hline
\hline
\multicolumn{8}{l}{$^{1}kT_{in}$ : Inner disk temperature,~~$^{2}kT_{e}$ : Temperature of the Compton cloud}\\
\multicolumn{8}{l}{$^{3}\tau$ : Optical depth of the Compton cloud,~~$^{4}\Gamma_{x}$ : Power-law photon index}\\
\multicolumn{8}{l}{$^{5}RQ$ : Radio-quiet,~~$^{6}RL$ : Radio-loud}\\
\end{tabular}
\end{center}
\end{table}

\setcounter{table}{4}
\begin{table*}
\caption{Spectral parameters during various classes of GRS 1915$+$105}
\begin{center}
\begin{tabular}{llllccc}
\hline
\hline
Model$^1$  &Normalization$^2$  &$\omega_{dip}$(A$^3$)  &$\beta_{low}$(C$^4$)  &$\theta_{low}$(A$^4$)  &$\phi$(A$^5$) &$\delta$(B$^6$)\\
\hline
\hline
$\chi2$ (RQ$^7$)   &dbb    &351.8  &118.5      &1605    &478.4       &2511\\
	       &co     &3.169  &7.835      &9.259   &1.624       &5.824\\
 &Reduced $\chi^2$    &9.96   &2.71       &24.09   &37.35       &173.7\\
\hline	       
\hline 	       
$\chi3$ (RL$^8$)   &dbb    &13.27  &3.6673E-10 &38.48   &9.710       &57.77\\
	       &co     &0.000  &4.8426E-33 &0.000   &0.000       &0.000\\
	       &po     &0.747  &13.97      &4.314   &2.0348E-33  &0.000 \\
  &Reduced $\chi^2$    &53.39  &8.41	   &27.66   &108.46      &127.3\\
\hline 	       
\hline 	       
$\omega$ (RQ)&dbb    &81.97    &4.6E-25   &316.2  &103.9       &316.2\\
(High state)    &co  &4.8E-14  &7.5E-25   &0.068  &0.000       &6.8E-02\\
	       &po   &10.0     &27.0      &28.63  &4.790       &28.63\\    
  &Reduced $\chi^2$  &0.84     &135.4 	  &2.02   &37.23       &2.02\\
\hline
\hline
\multicolumn{7}{l}{$^1$ : Model $\chi2$ : Model ``Disk-blackbody + CompST'' with
kT$_{in}$ = 1.348 keV,}\\
&\multicolumn{6}{l}{kT$_{e}$ = 20.01 and $\tau$ = 3.047}\\
\multicolumn{7}{l}{~~~~~Model $\chi3$ : ``Disk-blackbody + Power-law + CompST'' 
with kT$_{in}$ = 2.580 keV,}\\ 
&\multicolumn{6}{l}{kT$_{e}$ = 4.464, and $\tau$ = 46.56 and $\Gamma$ = 2.606}\\
\multicolumn{7}{l}{~~~~~Model $\omega$   : ``Disk-blackbody + Power-law + 
CompST'' with}\\ 
&\multicolumn{6}{l}{kT$_{in}$ = 1.709 keV, kT$_{e}$ = 2.532 keV and $\tau$ = 25.42 and $\Gamma$ = 2.912}\\
\multicolumn{7}{l}{$^2$ : dbb = Disk blackbody normalization, co = thermal-Compton spectrum }\\
&\multicolumn{6}{l}{normalization, and po = power-law normalization}\\
\multicolumn{7}{l}{$^3$ : Low-soft state of new class,~~~$^4$ : Low-hard state}\\
\multicolumn{7}{l}{$^5$ : Soft state,~~~$^6$ : High-soft state~~~$^7$ : Radio-quiet,~~~$^8$ : Radio-loud}\\
\end{tabular}
\end{center}
\end{table*}

\section{Discussion}

The X-ray observation of Galactic black hole candidates reveal four
different spectral states such as (a) {\it ``X-ray very high''} state 
with quite high soft X-ray flux and an ultra-soft thermal of multi-color 
blackbody spectrum of characteristic temperature {\it kT} $\sim$ 1 keV 
and a power-law tail with photon-index $\Gamma$ $\sim$ 2$-$3 with 
approximate X-ray luminosity at Eddington limit, (b) {\it ``X-ray high, 
soft''} state with similar characteristic temperature {\it kT} and a weak
power-law tail but with lower luminosity (by a factor of $\sim$ 3$-$30),
(c) {\it ``X-ray low, hard''} state with a single power-law spectrum with
photon-index $\Gamma$ $\sim$ 1.5$-$2 with a typical X-ray luminosity
of less than 1\% of Eddington, and (d) {\it ``X-ray off or quiescent''}
state with very low level emission with uncertain spectral shape at a 
luminosity {\it L$_X$ $<$ 10$^{-4}$} of Eddington limit (Grebenev et al. 
1993; van der Klis 1995). However, it is rare to observe a source exhibiting
all the four spectral states. 

\subsection{Fast transition between high-soft and low-hard states}

Galactic black hole binaries remain in a canonical spectral state with 
similar properties for considerably long durations ($\sim$ a few months). 
This suggests the general stable nature of the accretion disk. Though the 
microquasar GRS~1915$+$105 shows extended low-hard states as seen in other 
Galactic black hole candidates, the state transition between the low-hard 
state and high-soft state occurs in a wide range of time scales. Attempts 
have been made to explain the observed state transitions when the source 
shows regular periodic bursts in the X-ray light curves. Belloni et al. 
(1997) have tried to explain the repeated patterns (burst/quiescent cycle) 
in the X-ray light curve as the appearance and disappearance of the inner 
accretion disk. They have shown that the outburst duration is proportional 
to the duration of the previous quiescent state. Taam et al. (1997) have 
attempted to describe these transitions in the framework of thermal/viscous 
instabilities in the accretion disk. They have argued that the geometry of 
the accretion disk in GRS~1915$+$105 consists of a cold outer disk extending 
from radius $r_{in} \sim 30$ km to infinity and a hot, optically thin inner 
region between $r = 3$ and $r_{in}$. They have interpreted the spectral 
changes between low-hard and high-soft states as arising due to the change 
in the value of inner disk radius. Nayakshin et al. (2000) tried to explain 
the observed temporal behavior of GRS~1915$+$105 invoking the model of 
standard cold accretion disk with a corona that accounts for the strong 
nonthermal X-ray emission and plasma ejections in the jet when the source 
luminosity approaches the Eddington limit. This model qualitatively explains 
the observed cyclic features in the light curves (classes $\rho, \alpha, {\rm 
and} \lambda$), the dependence of the overall evolution and the values of the 
cycle times on the time-averaged luminosity, and the fact that the 
transitions between the states can be very much shorter than the 
corresponding cycle time. This model also successfully explains the 
ejections of plasma into radio jets and the associated dip features 
(class $\beta$) seen in the X-ray light curves of the source. 

Rao et al. (2000) have observed a slow transition from an extended low-hard 
state to a high-soft state ($\sim$ 3 months) in 1997 March$-$August. 
Fast transition (a few seconds) between the two spectral states is 
observed during many occasions when the source exhibits irregular bursts 
(Rao et al. 2000), soft dips (Naik et al. 2001) in the X-ray light curves. 
Chakrabarti et al. (2000) interpreted the observed spectral transition 
in GRS~1915$+$105 in the light of advective disk paradigm which includes 
self-consistent formation of shocks and out-flows from post-shock region. 
The observed fast transition between the two canonical spectral states 
implies the solutions for the accretion disk during two states exist for 
similar net (i.e., sum of the Keplerian and sub-Keplerian) $\dot{m}$. This 
is because the time scale of fast transition ($\sim$ 10 s) is not sufficient 
for the readjustment of the accretion disk at the outer edge to create a 
significant change in $\dot{m}$.

\subsection{Fast transition between high-soft and low-soft states}

The Galactic microquasar GRS~1915$+$105 remains in the low-hard state 
for extended periods and switches from the low-hard state into a high-soft
state in a wide range of time-scales. During the low-hard state, the source 
spectrum is dominated by the non-thermal component and the inner edge of 
the accretion disk lies far away from the black hole event horizon.
However, during the high-soft state, the spectrum is dominated by the thermal
component and the the inner edge of the accretion disk extends towards the
event horizon. The Compton cloud which is responsible for the Comptonization
of the soft X-ray photons during the low-hard state vanishes during the
high-soft state. The X-ray flux has a significant contribution from the 
non-thermal component during the low-hard state whereas the thermal component 
is dominated over the non-thermal component during the high intensity soft 
states. The transition between the above two canonical spectral states takes
place because of the infall of matter from the inner accretion disk into the
black hole and presence and absence of the Compton cloud. However,
the observed transition between the high-soft and low-soft states (present
work) without any significant change in the geometry of the accretion disk 
is interesting new phenomenon. We try to explain this observed feature  
invoking a model where the viscosity parameter of the accretion disk
is very close to the critical viscosity.

We observed a factor of $\sim$ 3.5 difference in the X-ray flux in 
2$-$60 keV energy range during the high-soft (non-dip) and low-soft (dip) 
states of class $\omega$. If this could be due to the decrease in the mass 
accretion rate, according to ADAF, the source spectrum during this low 
intensity state should be hard which is not the case. The softness of the 
source spectrum during the low state (dip) of class $\omega$ makes it clear 
that the observed change in intensity during these states (dip and non-dip) 
cannot be due to the change in mass accretion rate at the outer edge. Although 
similar change in X-ray flux during the dip and non-dip regions was observed 
in the light curves of the black hole candidates GRO~J1655$-$40 and 
4U~1630$-$47 (Kuulkers et al. 1998), these dips are different from those 
observed in GRS~1915$+$105. In earlier cases, the source spectra during the 
dips were heavily absorbed by some intervening material. The spectra during 
the dips when fitted by a model with power law and absorbed disk blackbody 
as the model components, the equivalent hydrogen column densities ($N_H$) 
for the two sources were found to be 27 $\times$ 10$^{22}$ cm$^{-2}$ and 
34 $\times$ 10$^{22}$ cm$^{-2}$ respectively which are about one order 
higher in magnitude than the interstellar absorption column densities 
($N_{H_{int}}$). The values of $N_H$ were very high ($\geq$ 76 $\times$ 
10$^{22}$ cm$^{-2}$) for both the sources when the spectra were fitted with 
a model with blackbody and absorbed disk blackbody as model components. 
However, the source spectra during the dips (low-soft state) in GRS~1915$+$105 
are well described by ``disk blackbody, power law and a thermal 
Compton-spectrum'' model without any absorbing medium other than $N_{H_{int}}$.
Hence, the decrease in X-ray intensity during the dips cannot be explained by 
absorption by the intervening medium. As the source is radio-quiet during the 
observations of this class, the decrease in X-ray intensity cannot be explained 
by the evacuation of matter from the accretion disk which causes flares in 
radio and infrared bands.

The observed unusual transition between two different intensity states
in GRS~1915$+$105 is attributed to the change in the temperature of the 
inner accretion disk without incorporating any significant change in the 
inner radius. The inner accretion disk during the high intensity state 
(non-dip) is hotter than the low intensity state (dip). The duration of
the observed dips and non-dips are in the range 20$-$95 s and 200$-$550 s.
If the variation in the intensity between two dips and non-dips are due
to the emptying and replenishing of the inner accretion disk caused by
a viscous thermal instability, then the viscous time can be explained by
(Belloni et al. 1997)
\begin{equation}
t_{vis} = 30 \alpha^{-1}_{2} M^{-1/2}_{1} R^{7/2}_{7} \dot{M}^{-2}_{18} s
\end{equation}
where $\alpha_{2} = \alpha/0.01, R_7$ is the radius in units of 10$^{7}$ cm,
M$_{1}$ is the mass of the compact object in solar masses, and $\dot{M}_{18}$ 
is the accretion rate in units of 10$^{18}$ g s$^{-1}$. Using all these 
parameters, Belloni et al. (1997) found that the model agrees with the data
with a relation of the form $t_{q} \propto R^{7/2}$, where $t_{q}$ is the
time interval for the quiescent phase. Applying the values of the radius of
the inner accretion disk during the dip to the above expression, the duration
of the derived quiescent (dip) period does not match with the observed
dip duration. These results manifest that the above model cannot explain the
the observed transition between two intensity states.  

We attempt to explain the observed phenomenon of state transition between 
different intensity states in GRS~1915$+$105 by invoking the two component 
advective flow (TCAF) model of Chakrabarti \& Titarchuk (1995) which consists
of two major disk components (i) standard, optically thick disk component
produced from the Keplerian or the sub-Keplerian matter at the outer 
boundary and (ii) quasi-spherical and axisymmetric sub-Keplerian halo 
component. The advantage of this model is that the soft and hard X-ray
radiations are formed self-consistently from the same accretion disk
without invoking any adhoc components such as the plasma cloud, hot corona,
etc. whose origins have never been clear. According to this model,
the temperature of the Keplerian disk increases with the increase in
the accretion rate of Keplerian disk. The increase in the number of soft 
photons intercepted by the post-shock region results in reducing its 
temperature. Assuming Comptonization as the dominant mechanism for cooling,
the expression for the electron temperature $T_e$ can be given as
\begin{equation}
T_e = \frac {T_{es} r_{s}}{r} e^{C_{comp}(r^{3/2}-r_{s}^{3/2})}
\end{equation}
where $C_{comp}$ is a monotonically increasing function of optical
depth assuming a constant spectral index, $r_{s}$ is the shock location 
and $T_{es}$ is the electron temperature at $r_{s}$. For $C_{comp}$ $>$ 
$r_{s}^{-3/2}$, the cooling due to Comptonization overcomes geometrical 
heating and $T_e$ drops as the flow approaches the  black hole. According 
to this model, a disk with completely free (Keplerian) disk and 
(sub-Keplerian) halo accretion rates for a black hole of mass $M = 5 M_\odot$, 
exhibits multiplicity in spectral index (when both the rates are fixed) or 
multiplicity in disk accretion rate when the spectral index is similar 
(see, Fig. 3a of Chakrabarti \& Titarchuk, 1995). In the 
observation described in this paper, there is no evidence for a large  
variation of the spectral index. This signifies that though there is not 
enough time for a change in the total rate (as the transition between the 
high-soft and low-soft states takes place within a time range of $\leq$ 10 s), 
individually, Keplerian and sub-Keplerian rates may have been modified. 
This is possible if the Shakura-Sunyaev viscosity parameter $\alpha$ is very 
close to the critical value ($\alpha \sim \alpha_c \sim 0.015$; Chakrabarti, 
1996). When $\alpha$ of the entire disk is well above $\alpha_c$, the entire 
disk is pretty much Keplerian, except very close to the black hole ($r<3r_g$, 
where $r_g$ is the Schwarzschild radii). Similarly when $\alpha$ of the entire 
flow is well below $\alpha_c$, the flow is sub-Keplerian with a possible 
standing or oscillating shock wave (Chakrabarti, 1996). Since viscosity in a 
disk can change in a very small time scale (convective/turbulent time-scale 
in the vertical direction) it is not unlikely that the viscosity near the 
Keplerian-disk surface is very close to the critical value during the time 
when this new class is exhibited. The high-intensity state will then 
correspond to the ordinary Keplerian disk. Extra-ordinary laminary flow may 
reduce viscosity at some stage, and sub-Keplerian flow develops out of the 
Keplerian disk  both above and below the Keplerian disk. This sub-Keplerian
flow need not be hotter since excess soft photons from the underlying 
Keplerian disk cools it instantaneously. Temperature and intensity of
the radiation from the Keplerian disk drops to the point that a thin 
outflow develops from the sub-Keplerian flow. This is the low intensity
state. This wind is cooled down and is fallen back on the Keplerian disk, 
increasing the Keplerian rate, intensity and viscosity, thereby cutting 
off the wind and bringing the flow to the high intensity state again. Unlike 
the transition from State B to State C  (as described by Belloni et al. 2000 
and explained in Chakrabarti et al. 2000), where the sub-Keplerian flow may 
always be present, in the present case, the high intensity state need not have 
a sub-Keplerian component at all. The transition from one soft-state to another
could in fact be due to the interesting change in topology of the flow at the
critical viscosity.

\section {Acknowledgments}
Work of SN is partially supported by the Kanwal Rekhi Scholarship of
the TIFR Endowment Fund. We thank the RXTE, BATSE, and NSF-NRAO-NASA GBI 
team for making the data publicly available. This research has made use of 
data obtained through the High Energy Astrophysics Science Archive Research 
Center Online Service, provided by the NASA/Goddard Space Flight Center. The 
Green Bank Interferometer is a facility of the National Science Foundation 
operated by the NRAO in support of NASA High Energy Astrophysics programs.

\end{document}